\documentclass[a4paper,10pt]{article}
\usepackage{graphicx}
\begin{document}

\twocolumn[%

\title{Transitions induced by the discreteness of molecules in a small autocatalytic system}
\author{Yuichi Togashi and Kunihiko Kaneko\\
{\small\it Department of Basic Science,
School of Arts and Sciences,
University of Tokyo,}\\
{\small\it Komaba, Meguro-ku, Tokyo 153-8902, Japan}}
\date{February 7, 2001}

\maketitle

\begin{abstract}
Autocatalytic reaction system with a
small number of molecules is studied numerically by stochastic
particle simulations. A novel state due to
fluctuation and discreteness in molecular numbers is found,
characterized as extinction of molecule species alternately
in the autocatalytic
reaction loop.  Phase transition to this state with
the change of the system size and flow is studied,
while a single-molecule switch of the molecule distributions is
reported.  Relevance of the results to intracellular processes are
briefly discussed.
\vspace{1mm}

\end{abstract}
\vspace{4mm}

]

Cellular activities are supported by biochemical reactions in a cell.
To study biochemical dynamic processes, rate equation for chemical reactions
are often adopted for the change of chemical concentrations.
However, the number of molecules in a cell
is often rather small \cite{Cell}, and it is not trivial if the rate equation
approach
based on the continuum limit is always justified.
For example, in cell transduction even a single molecule
can switch the biochemical state of a cell \cite{McAdams}.
In our visual system,
a single photon in retina is amplified to a macroscopic level \cite{retina}.

Of course, fluctuations due to a finite number of molecules is
discussed  by stochastic differential equation (SDE) adding
a noise term to the
rate equation for the concentration \cite{Kampen,Brussels}.
This noise term sometimes
introduces a non-trivial effect, as discussed as noise-induced phase
transition \cite{NIP},
noise-induced order \cite{NIO}, stochastic resonance \cite{SR}, and so
forth.  Still, these
studies assume that the average dynamics are governed by
the continuum limit, and the noise term is added as a perturbation to it.

In a cell, often the number of some molecules is very small, and may go down
very close to or equal to 0.  In this case, the change of the number
between zero and nonzero, together with the fluctuations
may cause a drastic effect that cannot be treated by SDE.
Possibility of some order different from macroscopic dissipative structure
is also discussed by Mikhailov and Hess \cite{Mikhailov1,Mikhailov2} (see
also Ref.\ \cite{Levine}).
Here we present a simple example with a phenomenon
intrinsic to a system with a small number of molecules
where both the fluctuations and digitality(`0/1') are essential.

In nonlinear dynamics, drastic effect of a single molecule may be expected
if a small change is amplified.
Indeed, autocatalytic reaction widely seen in
a cell, provides a candidate for such
amplification \cite{ES,Delbruck}.
Here we consider the simplest example of autocatalytic reaction
networks (loops) with a non-trivial finite-number effect.
With a cell in mind, we consider reaction of molecules
in a container, contacted with a reservoir of molecules.
The autocatalytic reaction loop is
\begin{math}
X_i + X_{i+1} \rightarrow 2X_{i+1}; i=1,\cdots,k; X_{k+1}\equiv X_1\\
\end{math}
within a container.
Through the contact with a reservoir,
each molecule $X_i$ diffuses in and out.

Assuming that the chemicals are well stirred in the container,
our system is characterized by the number of molecules $N_i$
of the chemical $X_i$ in the container with
the volume $V$ \cite{noteA}.
In the continuum limit with a large number of molecules,
the evolution of concentrations $x_i \equiv N_i/V$ is represented by
\begin{equation}
dx_i/dt=r_ix_{i-1}x_i-r_{i+1}x_ix_{i+1}+D_i(s_i-x_i)
\label{eqn:1}
\end{equation}
where $r_i$ is the reaction rate, $D_i$ is the diffusion rate across the surface of 
the container, and $s_i$ is the concentration of the molecule in the reservoir.

For simplicity, we consider the case $r_i=r$, $D_i=D$, and $s_i=s$ for all
$i$,
while the phenomena to be presented here will persist
by dropping this condition.
With this homogeneous parameter case, the above equation has a unique
attractor, a stable fixed point solution with $x_i=s$.
The Jacobi matrix around this fixed point
solution
has a complex eigenvalue, and the fluctuations around the fixed point
relax with the frequency $\omega_p \equiv rs/\pi$.
In the present paper we mainly discuss the case with $k=4$, since it is the
minimal number to see
the new phase to be presented.

If the number of molecules is finite but large, the reaction dynamics can be
replaced by Langevin equation by adding a noise term to eq.\ (\ref{eqn:1}).
In this case,
the concentration $x_i$ fluctuates around the fixed point,
with the dynamics of a component of the frequency $\omega_p$.
No remarkable change is observed with the increase of the noise strength,
that corresponds to the decrease of the total number of molecules.

To study if there is a phenomenon that is outside of this SDE approach,
we have directly simulated the above autocatalytic reaction model,
by colliding molecules stochastically.
Taking randomly a pair of particles
and examining if they can react or not, we have made the reaction with the
probability proportional to $r$.
On the other hand, the diffusion out to the reservoir is taken account of by
randomly sampling molecules and probabilistically removing them with
in proportion to the diffusion rate $D$, while the flow to the
container is also carried out stochastically in
proportion to $s$, $D$ and $V$ \cite{NOTE}.
Technically, we divide time into time interval $\delta t$
for computation, where one pair for the reaction, and single molecules
for diffusion in and out are checked.
The state of the container is always updated when a reaction
or a flow of a molecule has occurred.
The reaction $X_{i} + X_{i+1} \rightarrow 2X_{i+1}$ is made with the
probability $P_{Ri}(t,t+\delta t) \equiv r x_{i}(t) x_{i+1}(t) V \delta t
= r N_{i}(t) N_{i+1}(t) V^{-1} \delta t$ within the step $\delta t$.
A molecule diffuses out with the probability
$P_{Oi} \equiv DV x_{i}(t) = D N_{i}(t)$, and flows in with
$P_{Ii} \equiv DVs$.
We choose $\delta t$ small enough so that the numerical result
is insensitive with the further decrease of $\delta t$.
By decreasing $Vs$, we can control the average number of molecules in the
container,
and discuss the effect of a finite number of molecules, since
the average of the total number of molecules $N_{tot}$ is around the order
of $4Vs$ \cite{note1}.
On the other hand, the `discreteness' in
the diffusion is clearer as the diffusion rate $D$ is decreased.
We set $r=1$ and $s=1$, without loss of generality
($rs/D$ and $sV$ are the only relevant parameters of the model
by properly scaling the time $t$).

\begin{figure}
\includegraphics[width=\columnwidth]{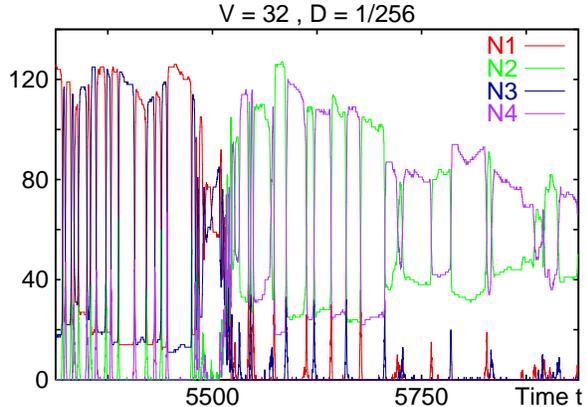}
\caption{(color)
Time series of the number of molecules $N_i(t)$,
for $D=1/256, V=32$.
Either 1-3 or 2-4 rich state is stabilized.
Successive switches appear
between $N_1 > N_3$ and $N_3 > N_1$ states
with $N_2, N_4 \approx 0$.
Here a switch from 1-3 rich to 2-4 rich state occurs
around $t=5500$.}
\label{fig:tseries}
\end{figure}

\begin{figure}
\includegraphics[width=\columnwidth]{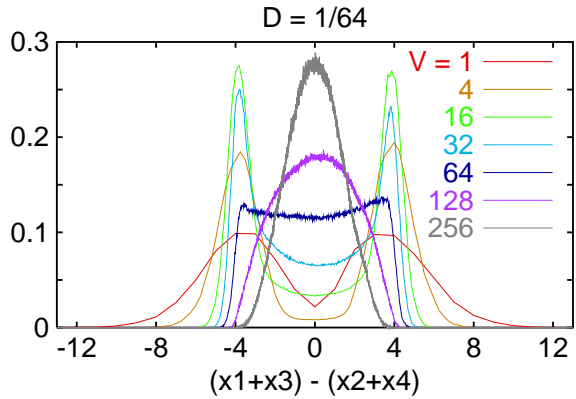}\\
\caption{(color)
The probability distribution of $z \equiv (x_1+x_3)-(x_2+x_4)$,
sampled over $2.1$ -- $5.2 \times 10^{6}$ steps. $D=1/64$.
For $V \ge 128$, $z$ has a distribution around 0, corresponding
to the fixed point state $x_i = s (= 1)$.
For $V \le 32$, the distribution has double peaks around
$z \approx \pm 4$,
corresponding to the state $N_1, N_3 \gg N_2, N_4 (\approx 0)$ or
the other way round.  The double-peak distribution is sharpest
around $V=16$, and with the further decrease of $V$, the distribution
is broader due to finite-size fluctuations.}
\label{fig:dist-ac-bd}
\end{figure}

First, our numerical results agree with those obtained by the
corresponding Langevin equation if $D$ and $V$ are not too small.
As the volume $V$ (and accordingly $N_{tot}$) is decreased, however, we
have found a new state whose correspondent does not exist
in the continuum limit.
An example of the time series is plotted in Fig.\ \ref{fig:tseries},
where we note
a novel state with $N_1,N_3 \gg 1$ and $N_2,N_4 \approx 0$
or $N_2,N_4 \gg 1$ and $N_1,N_3 \approx 0$.
To characterize this state quantitatively,
we have measured the probability distribution of $z \equiv
x_1+x_3-(x_2+x_4)$.  Since the
solution of the continuum limit is $x_i=s (=1)$ for all $i$,
this distribution has a sharp peak
around 0, with a Gaussian form approximately,
when $N_{tot}$is large enough.
As shown in Fig.\ \ref{fig:dist-ac-bd}, the
distribution starts to have double peaks around $\pm 4$,
as $V$ is decreased.
With the decrease of $V$ (i.e., $N_{tot}$),
these double peaks first
sharpen, and then
get broader with the further decrease
due to too large fluctuation of a system
with a small number of molecules.
Hence the new state with switches between 1-3 rich and 2-4 rich temporal
domains
is a characteristic phenomenon that appears only
within some range of a small number of molecules \cite{noteC}.

The stability of this state is understood as follows.
Consider the case with 1-3 rich and $N_2=N_4=0$.  When one (or few)
$X_2$ molecules flow in, $N_2$ increases, due to the autocatalytic reaction.
Then $X_3$ is amplified, and since
$N_2$ is not large, $N_2$ soon comes back to 0 again.  In short,
switch from $(N_1,0,N_3,0)$ to $(N_1-\Delta,0,N_3+\Delta+1,0)$
occurs with some $\Delta$, but the 1-3 rich state itself is maintained.
In the same manner, this state
is stable against the flow of $X_4$.
The 1-3 rich state is maintained unless
either $N_1$ or $N_3$ is close or equal to 0, and both
$X_2$ and $X_4$ molecules flow in within the switch time.  Hence the 1-3
rich state (as well as
2-4 rich state, of course) is stable as long as the flow rate
is small enough.

\begin{figure}
\includegraphics[width=\columnwidth]{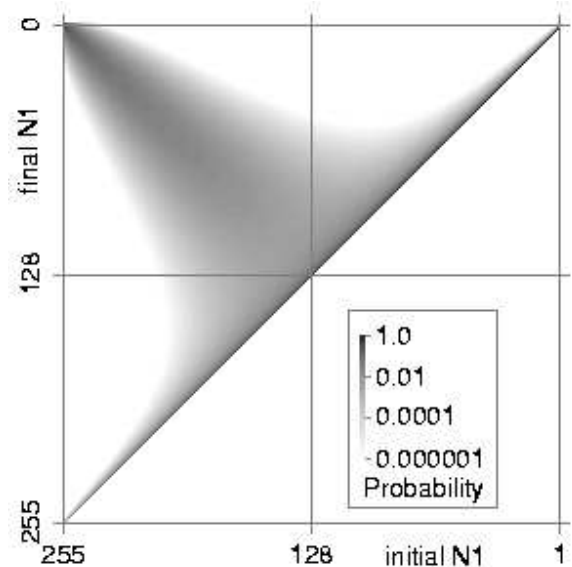}\\
\caption{Probability density for the switch
from $(N_1, N_3)$ to $(N_1', N_3')$
when a single $X_2$ molecule is injected into the system.
$N_1+N_3=N_{ini}$ is fixed at 256 initially.
There is no more flow and $N_4$ is always kept at 0,
so that the switch is completed when $N_2$ comes back to 0,
and $N_1'+N_3'=N_{ini}+1$.
Probability to take $N_1'$ is plotted against initial $N_1$.}
\label{fig:sw-prob}
\end{figure}

Within a temporal domain of 1-3 rich state, switches occur to change from
$(N_1, N_3) \rightarrow (N_1', N_3')$.
In Fig.\ \ref{fig:sw-prob},
we have plotted the probability density for the switch from $N_1
\rightarrow N_1' $
when a single $X_2$ molecule flows in, amplified, and $N_2$ comes back to 0,
by fixing $N_1+N_3=N_{ini}$ at
256 initially.
(We assume no more flow. Hence $N_1'+N_3'=N_{ini}+1$).
The peak around $N_1' \approx N_1 +1$ means the reaction from $N_2$ to
$N_3$ before the amplification, while
another peak around $N_1' \approx N_3=N_{ini}-N_1$ shows the conversion of
the
numbers through the
amplification of $X_2$ molecules.  Indeed, each temporal domain of
the 1-3 rich state consists of successive switches of $(N_1, N_3)
\rightarrow \approx (N_3,N_1)$,
as shown in Fig.\ \ref{fig:tseries}.
Since molecules diffuse out or in randomly besides this switch,
the difference between $N_1$ and $N_3$ is tended to decrease.  On
the other hand, each 1-3 rich state, when formed, has imbalance
between $N_1$ and $N_3$, i.e., $N_1 \gg N_3$ or $N_1 \ll N_3$,
since, as in Fig.\ \ref{fig:tseries},
the state is attracted from
alternate amplification of $X_i$, where only one type $i$ of molecules
has $N_i \gg 1$ and 0 for others. However,
the destruction of the 1-3 rich state is easier if $N_1 \gg N_3$ or
$N_1 \ll N_3$, as mentioned.
Roughly speaking, each 1-3 rich state starts with a large imbalance
between $N_1$ and $N_3$, and continues over a long time span,
if the switch and diffusion lead to $N_1 \approx N_3$, and
is destroyed when the large imbalance is restored. Indeed,
we have plotted the distribution of $y \equiv x_1-x_3+x_2-x_4$,
to see the imbalance for each 1-3 rich or 2-4 rich domain.
This distribution shows double peaks clearly around
$y \approx \pm 2.8, i.e., (N_1,N_3) \approx (3.4V,0.6V),(0.6V,3.4V)$.

Let us now discuss the condition to have the 1-3 or 2-4 rich state.
First, the total number of molecules should be
small enough so that the fluctuation from the state
$N_i\approx N_j$ (for $\forall i,j$)
may reach the state with $N_i \approx 0$.
On the other hand, if the total number is too small, even $N_1$ or $N_3$
for the 1-3 rich state may approach 0 easily, and the state is easily
destabilized. Hence the alternately rich state is stabilized only
within some range of $V$.

\begin{figure}
\includegraphics[width=\columnwidth]{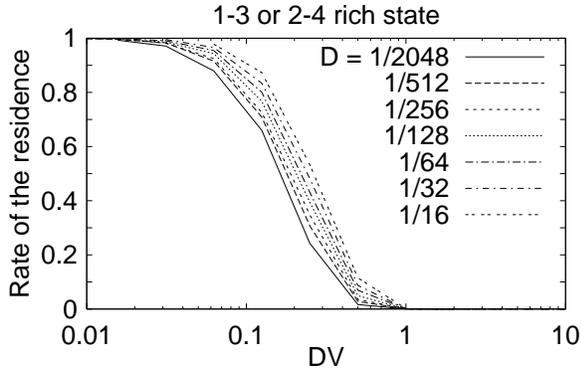}\\
\caption{The rate of the residence at 1-3 (or 2-4) rich state
over the whole temporal domain, plotted against $DV$ \cite{note2}.
Here, the residence rate is computed as follows.
As long as $N_2 > 0$ and $N_4 > 0$ are not satisfied simultaneously,
over a given time interval (8.0; 2.5 times
as long as the period of the oscillation around the fixed point
at continuum limit), it is
counted as the 1-3 rich state (2-4 rich state is
defined in the same way) \cite{note2a}.
The residence rate is computed as the ratio of the fraction of
the time intervals of 1-3 or 2-4 rich state to the whole interval.
For small $D$, this switching state is observed even for
a large number of molecules, say $N=4\times 10^{4}$, for
$V=10^{4}$ at $D=10^{-6}$.}
\label{fig:type1a}
\end{figure}

Note also that our system has conserved quantities
$\sum_i N_i$ (and $\sum_i log x_i$ in the continuum limit),
if $D$ is set at $0$.
Hence, as the diffusion rate gets
smaller, some characteristics of the
initial population are maintained over long time.  Once the above 1-3 (or
2-4) rich state
is formed, it is more difficult to be destabilized if $D$ is small.
In Fig.\ \ref{fig:type1a},
we have plotted the rate of the residence at 1-3 (or 2-4) rich
state over the whole temporal domain, with the change of $V$.
Roughly speaking, the state appears for $DV < 1$ \cite{note2},
while for too small $V$ (e.g., $V<4$),
it is again destabilized by fluctuations.
Although the range of the 1-3 rich state is larger for small $D$,
the necessary time to approach it increases linearly with $V$.
Hence it would be fair to state that properly small number of molecules
is necessary to have the present state.

To sum up, we have discovered a novel state in reaction dynamics
intrinsic to a small number of molecules.  This state is characterized by
alternately vanishing chemicals within an autocatalytic
loop, and switches by a flow of single molecules \cite{noteB}.
Hence, this state generally appears for a system with an autocatalytic loop
consisting of any even number of elements.
With the increase of $k$, however, the globally alternating state
all over the loop is more difficult to be reached.
In this case, locally alternating states are often formed with the
decrease of the system size (e.g., `2-4-6-8 rich' and `11-13-15 rich' states
for $k=16$).  This local order is more vulnerable to the flow of molecules
than the global order for the  $k=4$ loop.

On the other hand, for $k=3$, two of the chemical species
start to vanish for small $V$, since
any pair of different chemical species can react
so that one chemical species is quickly absorbed into the other.
This state of single chemical species, however,
is not stable by a flow of a single molecule.
Indeed, no clear `phase transition'
is observed with the decrease of $V$.

Although in the present Letter we have studied the case with $s_i=s$,
we have also confirmed that the present
state with alternately vanishing chemical species is generally
stabilized for small $V$, even if
$s_i$ or $r_i$ or $D_i$ are not identical.

Last, we make a remark about the signal transduction in a cell.
In a cell, often the number of molecules is small, and the cellular
states often switch by a stimulus of a single molecule \cite{Cell}.
Furthermore,
signal transduction pathways generally include autocatalytic reactions.
In this sense, the present stabilization of the alternately rich state
as well as
a single-molecule switch may be relevant to cellular dynamics.
Of course, one may wonder that the present mechanism is
too `stochastic'. Then, use of both the present mechanism
and robustness by dynamical systems \cite{KKYOMO,CFKK} may be important.
Indeed, we have made some preliminary simulations of
complex reaction networks.
Often, we have found the transition to a new state at a small number
of molecules, when the network includes the autocatalytic loop of
4 chemicals as studied here \cite{note3}.
Hence the state presented here is not restricted to this specific
reaction network, but is observed in a class of autocatalytic
reaction network.
Furthermore switches between different dynamic states (limit cycles or
chaos)
are possible when the number of some molecules (that are not directly
responsible to the switch)
is large enough.  The `switch of dynamical systems' by the
present few-number-molecule mechanism
will be an important topic to be pursued in future.

We would like to thank C. Furusawa, T. Shibata and T. Yomo
for stimulating discussions.
This research was supported by
Grants-in-Aid for Scientific
Research from the Ministry of Education, Science, and Culture
of Japan (Komaba Complex Systems Life Science Project).



\begin{thebibliography}{99}
\bibitem{noteA}
In a cell, reaction often takes place at a localized part
of a cell. In this case,
the volume $V$ in our model does not necessarily represent the whole cell
volume, but the volume of such localized region.
\bibitem{NOTE}
One might assume the choice of the diffusion flow proportional to
$DV^{2/3}$, considering the area of surface.  Here we choose the
flow proportional to $DV$, to have a well-defined continuum limit (eq.(1))
for $V \rightarrow \infty$.  At any rate, by just re-scaling $D$, the
present
model can be rewritten into the case with $DV^{2/3}$, for finite $V$.
Hence the result here is valid for the $DV^{2/3}$ (and other) cases.
\bibitem{note1}
For small $V$ value,
there appears deviation from this estimate.
At any rate, the average number decreases monotonically with $V$, and for
$V=0$, it goes to zero.
\bibitem{noteC}
One may assume that a similar switch could exist
in an equilibrium system due to a finite-size effect,
e.g., in a small magnetic system with two ordered 
states of spins up and down. There, the two ordered states
exist in the continuum (thermodynamic) limit.
In contrast, the present `1-3 rich' and `2-4 rich' states
to be switched do not exist in the continuum limit, but appear
only through the discreteness of the number of molecules
(and non-equilibrium dynamics allowing for oscillatory relaxation).
\bibitem{note2}
As shown in Fig.\ \ref{fig:type1a},
there is a deviation from the scaling by $DV$.
All the data are fit much better either by $D^{0.9}V$ or by $(D+0.0002)V$.
At the moment we have no theory which form is justified.
\bibitem{noteB}
The present transition does not require any details of a cell, but
only chemical reactions within a small system surrounded by a
permeable membrane. Hence, it can be experimentally verified
by studying autocatalytic chemical reactions within a
very small container, for example, within
an artificially synthesized liposome.
\bibitem{note2a}
This estimate includes the case in which only one species exists,
and gives an overestimate for very small V.
\end{thebibliography}
\end{document}